\documentclass[aps,showpacs,preprintnumbers,amsmath,amssymb,prb,twocolumn,superscriptaddress
]{revtex4-1}
\usepackage{txfonts}
\usepackage{graphicx}
\usepackage{dcolumn}
\usepackage{bm}
\usepackage{color}
\usepackage{subfigure}  
\def\comment#1{}

\def\slashchar#1{\setbox0=\hbox{$#1$}           
   \dimen0=\wd0                                 
   \setbox1=\hbox{/} \dimen1=\wd1               
   \ifdim\dimen0>\dimen1                        
      \rlap{\hbox to \dimen0{\hfil/\hfil}}      
      #1                                        
   \else                                        
      \rlap{\hbox to \dimen1{\hfil$#1$\hfil}}   
      /                                         
   \fi}                                         %

\def\sigmab{{\mbox{\boldmath $\sigma$}}}
\def\nablab{{\mbox{\boldmath $\nabla$}}}

\def\gammab{{\mbox{\boldmath $\gamma$}}}

\begin{document}

\title{Thermal screening at finite chemical potential
on a topological surface and its interplay with proximity-induced ferromagnetism}

\author{Flavio S. Nogueira}
\affiliation{Institut f{\"u}r Theoretische Physik III, Ruhr-Universit\"at Bochum,
Universit\"atsstra\ss e 150, DE-44801 Bochum, Germany}

\author{Ilya Eremin}
\affiliation{Institut f{\"u}r Theoretische Physik III, Ruhr-Universit\"at Bochum,
Universit\"atsstra\ss e 150, DE-44801 Bochum, Germany}
\affiliation{National University of Science and Technology “MISiS”, Moscow, 119049, Russian Fedeation}

\date{Received \today}

\begin{abstract}
Motivated by recent experiments on EuS/Bi$_2$Se$_3$ heterostructures, we study the temperature
dependent screening effects on the surface
of a three-dimensional topological insulator proximate to a ferromagnetically ordered system. In general, we find that besides
the chemical potential and temperature, the screening energy scale also depends on
the proximity-induced electronic gap in an essential way. In particular, at zero temperature the screening energy
vanishes if the chemical potential is smaller than the proximity-induced electronic gap. We show that  at finite temperature, $T$, and/or
chemical potential, $\mu$, the Chern-Simons (topological)
mass, which is generated by quantum fluctuations arising from the proximity-effect, can be calculated
analytically in the insulating regime. In this case the topological mass yields the Hall conductivity associated to edge states.
We show that when the chemical potential is inside the gap the topological mass remains nearly quantized at finite temperature.
\end{abstract}

\maketitle

\section{Introduction}

Recently three-dimensional topological insulators (TIs) with magnetically ordered surface states had attracted much attention,
both theoretically
\cite{Nagaosa-2010,Nagaosa-2010-1,Zang-Nagaosa,Franz-2010,Rosenberg-2010,Belzig-2012,Loss-PRL-2012,Nogueira-Eremin-2012,Cortijo-2012,Schmidt-2012,Qi-2013,Chulkov}
and experimentally.\cite{Hor,Wray,Vobornik,Rader-2012,Checkelsky,Moodera-2012,Moodera,kapitulnik-2013} The
topological surface states, usually gapless and featuring a low-energy Dirac dispersion protected by time-reversal (TR) symmetry,
generally become gapped when a ferromagnetically ordered layered material is grown over it. In this case the TR symmetry is broken by
the proximity effect to the magnetic material. Interesting effects on
the magnetization dynamics arise when the topological surface
becomes ferromagnetic.\cite{Nagaosa-2010,Franz-2010,Loss-PRL-2012,Nogueira-Eremin-2012} Quantum fluctuations of the
surface Dirac fermions induce an additional
Berry phase, which modifies the Landau-Lifshitz (LL) equation for the magnetization dynamics.\cite{Nagaosa-2010,Nogueira-Eremin-2012}
Furthermore, when an electric field is present, either external or originating from Coulomb
interactions, a topological magnetoelectric torque also appears in the LL equation.\cite{Nogueira-Eremin-2012} Actually
both effects arise from an induced Chern-Simons (CS) term \cite{CS} upon integrating out the Dirac fermions.\cite{Nogueira-Eremin-2012,Nogueira-Eremin-2013}

It is important to notice that most of the theoretical calculations in TIs were performed so far only at zero temperature,
although some aspects like the shift of the Curie temperature in the ferromagnetic insulator/TI
heterostructure were discussed.\cite{Nogueira-Eremin-2012,Moodera}
The disregard of the temperature effects till now is partially connected to the fact that finite temperature is not expected to modify
the topological contribution to the effective
electromagnetic Lagrangian of a TI in the bulk,\cite{Qi-2008}
\begin{equation}
\label{Eq:L-EM-TI}
 {\cal L}_{\rm EM}=\frac{1}{8\pi}\left(\epsilon{\bf E}^2-\frac{1}{\mu_0}{\bf B}^2\right)
+\frac{\alpha}{4\pi^2}\theta~{\bf E}\cdot{\bf B},
\end{equation}
where $\alpha=e^2/(\hbar c)$ is the fine-structure constant and $\theta$ is the so-called axion field \cite{Axions}.
While both the dielectric constant, $\epsilon$, and magnetic permeability, $\mu_0$, receive finite temperature corrections,
$\theta$ remains temperature-independent.\cite{Liu-Ni-PRD-1988} This somewhat confirms the naive expectation for the term being topological in origin.
This, however, is not the full story.  For example, differently from the chiral anomaly
leading to the axion term in Eq. (\ref{Eq:L-EM-TI}), the parity anomaly
in $d=2+1$ dimensions does indeed receive finite temperature corrections.\cite{CS-finite-T} Therefore, integrating out the
Dirac fermions in 2+1 dimensions and at finite temperature generates additional temperature-dependent CS term.

Another important finite temperature aspect relates to the electromagnetic response and
the many-body screening of the Coulomb interaction.
In graphene, where the chemical potential is typically zero, there is no
screening of the long-range Coulomb interaction
at zero temperature, although a renormalization of the dielectric constant occurs.\cite{interac-graphene}
However, at finite temperature a thermal screening should occur, corresponding to a
situation reminiscent of massless QED in $2+1$ dimensions.\cite{Dorey-1991} Moreover, in contrast to graphene, most
three-dimensional TIs feature surface Dirac fermions at a non-vanishing chemical potential, and we may expect
that the screening in TIs \cite{Zang-Nagaosa,Screen-TI}
and in doped graphene \cite{Ando-2006} at zero temperature behaves differently.

In recent experiments \cite{Moodera,kapitulnik-2013}
thin films of EuS/Bi$_2$Se$_3$ heterostructure were grown, and it has been shown that the topological surface
becomes ferromagnetic due to a proximity-induced symmetry breaking mechanism. The experiments also confirmed
the theoretical expectations that the surface Dirac fermions become gapped.\cite{Moodera,kapitulnik-2013}
Many measurements in such a heterostructure are made at temperatures close to
the Curie temperature. Therefore, it is of paramount importance to study theoretically quantum field theory
models of Dirac fermions at finite temperature and chemical potential. As we have mentioned above, there are finite temperature corrections
to the generated CS term. Furthermore, since the CS term provides a correction to the Berry phase of the proximity-induced
ferromagnetism,\cite{Nagaosa-2010,Nogueira-Eremin-2012} the precessional behavior of
the magnetization will be modified at finite temperature and in turn affect the spin-wave excitations.

A finite chemical potential influences the physics on the surface dramatically even at zero temperature. For example, it is known
that when the chemical potential is larger than the gap, the Hall conductivity is not quantized any longer and is given
by \cite{Zang-Nagaosa} $\sigma_{xy}=e^2m_0/(2h\mu)$. Here $m_0$ represents the zero temperature gap. This suggests that
in such a metallic regime the Hall conductivity cannot always be identified with the coefficient of a
{\it local} fluctuation-generated CS term, i.e.,
the topological mass which is induced by quantum fluctuations rather than by an external response.
Since the relevant situation for the
magnetization dynamics is when the chemical potential is inside the gap, it is important to ask whether the Hall conductivity
remains quantized in this case when the temperature is finite.
In this paper we will derive an analytical expression for Hall conductivity at
finite temperature and chemical potential in precisely this regime. We will show that a plateau persists at finite temperature
when the chemical potential is inside the gap.

The plan of the paper is as follows. In Section II we introduce the model and discuss some simple thermally induced screening effects following
from the vacuum polarization on the TI surface. In Section III we derive the low-energy CS term at finite temperature and
chemical potential.
We discuss the physical consequences of
this result for the magnetization dynamics in Section IV, and in Section V we summarize the main results.

\section{Model and vacuum polarization at finite temperature}

Our starting point is the following Hamiltonian for the surface Dirac fermions,
%
 $H=\psi^\dagger {\cal H}~\psi$,
where $\psi=[\psi_\uparrow, \psi_\downarrow]^T$. The $2\times 2$ matrix ${\cal H}$ reads,
\begin{equation}
\label{Eq:matrix-H}
 {\cal H}=v_F(-i\hbar\nablab\times\hat {\bf z})\cdot{\sigmab}-e\varphi-J(n_x\sigma_x+n_y\sigma_y)-J_\perp n_z\sigma_z,
\end{equation}
where $\varphi$ is a scalar potential including contributions
both from external electric field and  Coulomb interaction, and
the magnetization ${\bf n}$ satisfies the constraint ${\bf n}^2=1$. The Lagrangian
${\cal L}=\psi^\dagger i\hbar\partial_t\psi-H$ can be after a rescaling
$\varphi\to (J/e)\varphi$ conveniently written in QED-like form,\cite{Nogueira-Eremin-2012,Nogueira-Eremin-2013}
\begin{equation}
 \label{Eq:TI-QED}
 {\cal L}=\bar \psi(i\slashchar{\partial}-J\slashchar{a}-J_\perp n_z)\psi,
\end{equation}
In the above equation the Dirac slash notation is used, $\slashchar{Q}=\gamma^\mu Q_\mu$, with
$\gamma^0=\sigma_z$, $\gamma^1=-i\sigma_x$, and $\gamma^2=i\sigma_y$, and $\partial_\mu=\hbar(\partial_t,v_F\nablab)$,
$\bar \psi=\psi^\dagger\gamma^0$, and the
vector field $a^\mu=(a_0,{\bf a})=(\varphi,n_y,-n_x)$.

In order to study thermal effects on the magnetization dynamics and the screening of the Coulomb interaction,
we will perform the calculations in the imaginary time formalism, setting as
usual $t=-i\tau$ with $\tau\in[0,\beta]$, and integrating out the fermions with anti-periodic boundary conditions,
$\psi(0,{\bf r})=-\psi(\beta,{\bf r})$. Thus, the amplitude $\exp(i\hbar^{-1}\int d^3x{\cal L}_F)$ appearing in the functional integral
becomes $\exp(-\hbar^{-1}\int_0^\beta d\tau\int d^2r{\cal L}_F^{\rm eucl})$
in the imaginary time formalism, with the Lagrangian in euclidean spacetime given by,
\begin{equation}
 \label{Eq:TI-QED-eucl}
 {\cal L}^{\rm eucl}=\bar \psi(\slashchar{\partial}-\mu\gamma_0-iJ\slashchar{a}+J_\perp n_z)\psi,
\end{equation}
where now the Dirac slash features the euclidean Dirac matrices $\gamma_\mu=(\sigma_z,\sigma_x,-\sigma_y)$ and
a chemical potential $\mu$ was included (note that $\psi^\dagger\psi=\bar \psi\gamma_0\psi$).
Integrating out the fermions yields the effective action in the form,
%
 ${\cal S}_{\rm eff}={\cal S}_F+{\cal S}_{\rm mag}({\bf n})$,
%
with
\begin{equation}
 \label{Eq:Sf}
 {\cal S}_F=-\frac{N}{V}{\rm Tr}\ln(\slashchar{\partial}-\mu\gamma_0-iJ\slashchar{a}+J_\perp n_z),
\end{equation}
where $N$ is the number of Dirac fermion species and $V$ is the (infinite) volume.
${\cal S}_{\rm mag}({\bf n})$ is the magnetic action, which has the general form in the imaginary time formalism,
\begin{equation}
 \label{Eq:S-mag}
 {\cal S}_{\rm mag}=\int d\tau\int d^2r\left[{\bf b}({\bf n})\cdot\partial_\tau{\bf n}+{\cal H}_{\rm mag}\right],
\end{equation}
featuring a Berry gauge potential satisfying $\nablab_{\bf n}\times{\bf b}={\bf n}$, representing a
magnetic monopole in the magnetization space. The magnetic Hamiltonian
density ${\cal H}_{\rm mag}$
may contain several contributions, the most important ones being the coupling to external fields and the exchange terms.

In TIs the number $N$ of fermion species is odd. In the present context this is essential, otherwise
no parity and TR symmetry breaking via the generation of a CS term would occur dynamically.\cite{Nogueira-Eremin-2013}
Assuming that the system orders along the $z$-axis, we can write $n_z=\langle n_z\rangle+\tilde n_z$ and expand
Eq. (\ref{Eq:Sf}) up to quadratic order in the fields,
%
\begin{eqnarray}
 &&{\cal S}_F\approx\frac{N}{2}\int_0^\beta d\tau\int d^2r\int_0^\beta d\tau'\int d^2r'
 \left[\Pi_{\mu\nu}(\tau-\tau',{\bf r}-{\bf r}')\right.\nonumber\\
 &\times&\left.a_\mu(\tau,{\bf r})a_{\nu}(\tau',{\bf r}')
 +\chi_{zz}(\tau-\tau',{\bf r}-{\bf r}')\tilde n_z(\tau,{\bf r})\tilde n_z
 (\tau',{\bf r}')
 \right],
\end{eqnarray}
where
%
 $\Pi_{\mu\nu}(\tau-\tau',{\bf r}-{\bf r}')=\delta^2{\cal S}_F/[\delta a_\mu(\tau,{\bf r})
 \delta a_\nu(\tau',{\bf r}')]|_{n_z=\langle n_z\rangle}$,
is the vacuum polarization tensor at finite temperature encompassing screening effects in
the Coulomb potential and the transverse magnetic susceptibility, and
%
 $\chi_{zz}(\tau-\tau',{\bf r}-{\bf r}')=\delta^2{\cal S}_F/[\delta \tilde n_z(\tau,{\bf r})
 \delta \tilde n_z(\tau',{\bf r}')]|_{n_z=\langle n_z\rangle}$,
is the longitudinal magnetic susceptibility.
The fermionic propagator has the form, $G_F=(\slashchar{\partial}-\mu\gamma_0+m)^{-1}$, or in momentum space,
$G_F(p)=(i\slashchar{p}+\mu\gamma_0+m)^{-1}=[m-(i\omega_n+\mu)\gamma_0-i\gammab\cdot{\bf p}]/[m^2+{\bf p}^2-(i\omega_n+\mu)^2]$, where
$p=(p_\mu)=(p_0,{\bf p})=(\omega_n,v_F{\bf k})$, $m=J_\perp\langle n_z\rangle$, and
$\omega_n=(2n+1)\pi/\beta$ is the usual fermionic Matsubara
frequency. The units are such that $\hbar=1$.

The calculation of the vacuum polarization in $2+1$ dimensions and zero temperature is well known \cite{CS}
and has been reviewed by us in detail recently in Ref. \onlinecite{Nogueira-Eremin-2013}. Due to current conservation, it fulfills
$p_\mu\Pi_{\mu\nu}=0$. Periodicity in the Matsubara time allows
one to choose a rest frame for the heat bath given by the vector $u_\mu=(1,0,0)$.\cite{Kapusta}
Therefore, in this case we can write the vacuum polarization in momentum space in the form,
\begin{equation}
\label{Eq:Pi-decomp}
 \Pi_{\mu\nu}(p)=A(p)P_{\mu\nu}^T+B(p)P_{\mu\nu}^L+C(p)\epsilon_{\mu\nu\lambda}p_\lambda,
\end{equation}
where $P_{\mu\nu}^T$ and $P_{\mu\nu}^L$ are both transverse in 2+1 dimensions, with
$P_{\mu\nu}^T$ being transverse and $P_{\mu\nu}^L$ longitudinal in two spatial dimensions. Thus, we have
$P_{0i}^T=P_{00}^T=0$, $P_{ij}^T=\delta_{ij}-p_ip_j/{\bf p}^2$, and $P_{\mu\nu}^T+P_{\mu\nu}^L=\delta_{\mu\nu}-p_\mu p_\nu/p^2$,
where Latin indices refer to spatial dimensions.  At the same time,
%
\begin{eqnarray}
 \label{Eq:v-pol}
 &&\Pi_{\mu\nu}(\omega_n,{\bf p})=-NJ^2T
 \nonumber\\
 &\times&\sum_{l}\int\frac{d^2k}{(2\pi)^2}
 {\rm tr}[\gamma_\mu G_F(\nu_l,{\bf k})\gamma_\nu G_F(\nu_l+\omega_n,{\bf k}+{\bf p})],
\end{eqnarray}
where $\nu_l=\pi(2l+1)T$ and $\omega_n=2\pi nT$, $l,n\in\mathbb{Z}$.

At finite temperatures, an interesting aspect of the vacuum polarization in relativistic-like fermionic systems is the
generation of a thermal mass for the vector field along the time direction.\cite{Kapusta,Dorey-1991}
In other words, the Coulomb potential acquires a thermal gap or thermal screening.
In the case of a TI surface, for example, the Coulomb interaction $\phi_C(r)=e^2/(\epsilon r)$  in momentum space is given
by $\phi_C({\bf q})=2\pi e^2/(\epsilon|{\bf q}|)$,
similarly to interacting graphene.\cite{interac-graphene} The vacuum polarization screens this Coulomb interaction, and we have,
%
 $\phi_{\rm eff}({\bf q})=J^2/\{J^2[\phi_c({\bf q})]^{-1}+\Pi_{00}(0,{\bf q})\}$,
allowing us to define the so called
{\it electric mass},\cite{Kapusta} $m_{\rm el}^2\equiv\Pi_{00}(0,0)$.
An explicit calculation assuming $|{\bf q}|\ll \sqrt{2(\mu^2-m^2)}$ yields,
\begin{widetext}
\begin{equation}
\Pi_{00}(0,{\bf q})=\frac{NJ^2}{2\pi v_F^2}\left\{
 T\ln[\cosh(|m|/T)+\cosh(\mu/T)]
 -\frac{{\bf q}^2+2m^2}{\sqrt{{\bf q}^2+4m^2}}\frac{\sinh\left(\frac{\sqrt{{\bf q}^2+4m^2}}{2T}\right)}{\cosh(\mu/T)
 +\cosh\left(\frac{\sqrt{{\bf q}^2+4m^2}}{2T}\right)}\right\},
\end{equation}
 \end{widetext}
so that the electric mass is given by,
\begin{eqnarray}
 \label{Eq:m-el}
 m^2_{\rm el}&=&\frac{NJ^2}{2\pi v_F^2}\left\{\vphantom{\frac{1}{2}}T\ln[\cosh(|m|/T)+\cosh(\mu/T)]
\right.\nonumber\\
&-&\left.\frac{|m|\sinh(|m|/T)}{\cosh(|m|/T)+\cosh(\mu/T)}\right\}.
\end{eqnarray}
%
At zero temperature, $m_{\rm el}^2|_{T=0}=[NJ^2\mu/(2\pi v_F^2)]\theta(\mu-|m_0|)$, where $m_0=\lim_{T\to 0}m(T)$. Thus,
the electric mass vanishes for $0\leq \mu<|m_0|$.
Using the above results, we obtain that in the long wavelength limit the effective Coulomb interaction becomes,
\begin{equation}
 \label{Eq:Veff}
 \phi_{\rm eff}({\bf q})\approx \frac{J^2}{J^2[\phi_c({\bf q})]^{-1}+\Pi_{00}(0,0)}=\frac{2\pi e^2}{\epsilon(|{\bf q}|+s)},
\end{equation}
where $s=2\pi e^2m_{\rm el}^2/(\epsilon J^2)$, featuring in this way a screening of the Thomas-Fermi type.
At zero temperature, $s|_{T=0}\equiv s_0=Ne^2\mu\theta(\mu-|m_0|)/(\epsilon v_F^2)$.
In real space we have,
\begin{equation}
\label{Eq:Veff-real}
 \phi_{\rm eff}(r)=\frac{e^2}{\epsilon r}\{1+(\pi sr/2)[Y_0(sr)-H_0(sr)]\},
\end{equation}
where $Y_0(x)$ is a Bessel function of second kind and $H_0(x)$ is a Struve function.

The above results imply that
no screening would occur at zero temperature if $\mu<|m_0|=J_\perp|\langle n_z\rangle|$, i.e., the screening takes
place only when the chemical potential is located above the gap. This is a reasonable result, since this regime corresponds
to a metallic state.

For temperatures
above the Curie temperature, $T_c$, we have,
\begin{equation}
\label{Eq:s-large-T}
 s|_{T\geq T_c}=\frac{Ne^2}{\epsilon v_F^2}T\ln\left[2\cosh^2\left(\frac{\mu}{2T}\right)\right].
\end{equation}
The above equation is also valid when no ferromagnetic material is in contact with the TI surface. In
this case Eq. (\ref{Eq:s-large-T}) is valid for all $T\geq 0$. Interestingly, since we are dealing with
interacting Dirac fermions, we have that in absence of ferromagnetism Eq. (\ref{Eq:s-large-T}) would also
apply to doped graphene \cite{Ando-2006}
if one replaces $N\to N/2$ to account for the even number of Dirac cones.

\section{The Chern-Simons term and the Hall conductivity at finite temperature}

Next, we consider the finite temperature behavior of the CS term, which will yield a finite temperature correction
to the Berry phase on the TI surface.
The finite temperature behavior of the CS term is generally a highly
nontrivial matter,\cite{CS-finite-T,Dunne-1997} depending on the type of background field configuration involved.
However, in our case we are mainly interested in the low-energy behavior, so that the calculation simplifies
considerably.
The calculation parallels the zero temperature one,\cite{Nogueira-Eremin-2013}
and thus $C(p)=-2NJ^2mI(p)$, where
\begin{eqnarray}
 \label{Eq:Integral}
 I(\omega_n,{\bf p})&=&
 T\sum_l\int\frac{d^2 k}{(2\pi)^2}\frac{1}{(i\nu_l+\mu)^2-v_F^2{\bf k}^2-m^2}
 \nonumber\\
 &\times&\frac{1}{(i\nu_l+i\omega_n+\mu)^2-v_F^2({\bf k}-{\bf p})^2-m^2}.
\end{eqnarray}
When the chemical potential is inside the gap, i.e., $\mu<|m|$,
the surface state is insulating and we can take
$I(p)\approx I(0)$ in the low-energy regime, since in this case $m$ yields the dominant energy scale in the problem and
there is no Fermi momentum.
The effective action is local in this case, following straightforwardly from  a derivative expansion.
On the other hand, in the metallic regime where $\mu>|m|$, the effective action is inherently non-local. In this case there is
a Fermi surface singularity with a Fermi momentum $p_F=\sqrt{\mu^2-m^2}$.

Generally, the coefficient of the CS term, the so called topological mass, yields a measure of the Hall conductivity, when
scaled in appropriate unities. We will now show that
in the insulating regime, the topological mass can be evaluated analytically at finite temperature and chemical potential.
In this regime the CS contribution to the effective
action is given approximately by,
\begin{eqnarray}
\label{Eq:CS-action}
 {\cal S}_{\rm CS}\approx\frac{\sigma(T,m)
 }{2}\int dt\int d^2r \epsilon_{\mu\nu\lambda}a^\mu\partial^\nu a^\lambda
\end{eqnarray}
where in the above expression we have returned to real time and have defined $\sigma(T,m)=2NJ^2I(0)m$.
Expressing $\sigma(T,m)$ in units of conductivity via $\sigma(T,m)=NJ^2\tilde \sigma(T,m)/(v_F^2e^2)$ and using that
 $I(0)=[f_-(0)-f_+(0)]/(8\pi|m|)$ (see appendix A), we obtain,
\begin{equation}
\label{Eq:sigma}
 \tilde \sigma(T,m)=\frac{e^2{\rm sgn}(m)\sinh(|m|/T)}{2h[\cosh(|m|/T)+\cosh(\mu/T)]},
\end{equation}
where we have reintroduced the Planck constant.
The above equation provides an analytical expression for
the Hall conductivity for the case where the chemical potential is inside the gap, i.e., in
the insulating regime. However, it differs from
the actual Hall conductivity $\sigma_{xy}$
when $\mu>m$. In order to see this, let us first compare Eq. (\ref{Eq:sigma}) to the Hall
conductivity at zero temperature, where it can be calculated exactly for a nonzero chemical potential.
\begin{figure}[h!]
\centering
\subfigure[]{
\includegraphics[scale=0.32]{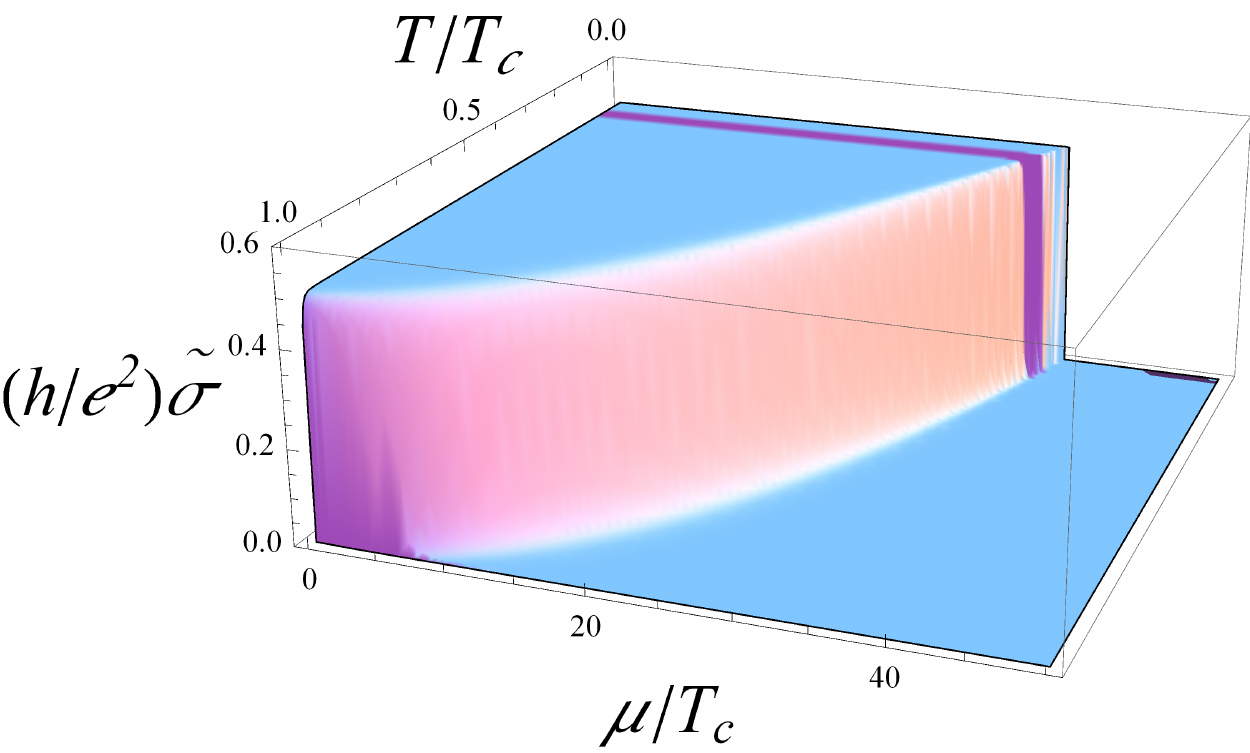}}
~
\subfigure[]{
\includegraphics[scale=0.32]{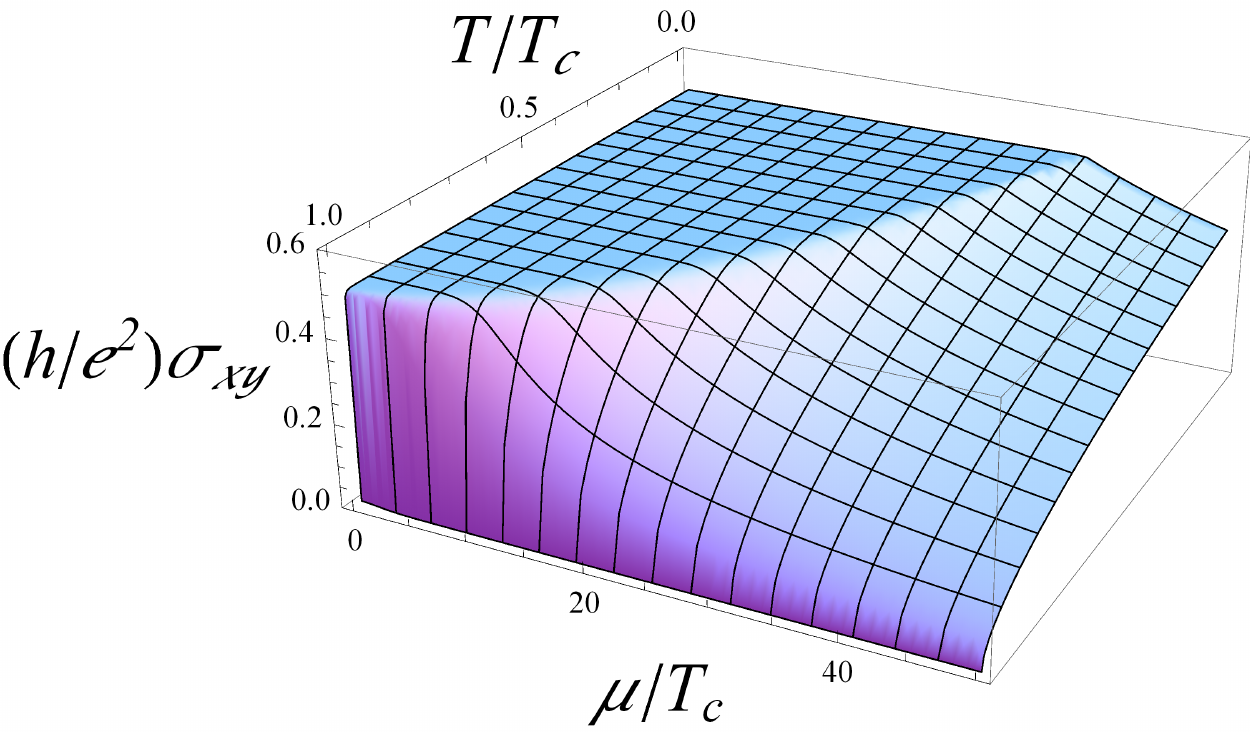}}

\subfigure[]{
\includegraphics[scale=0.32]{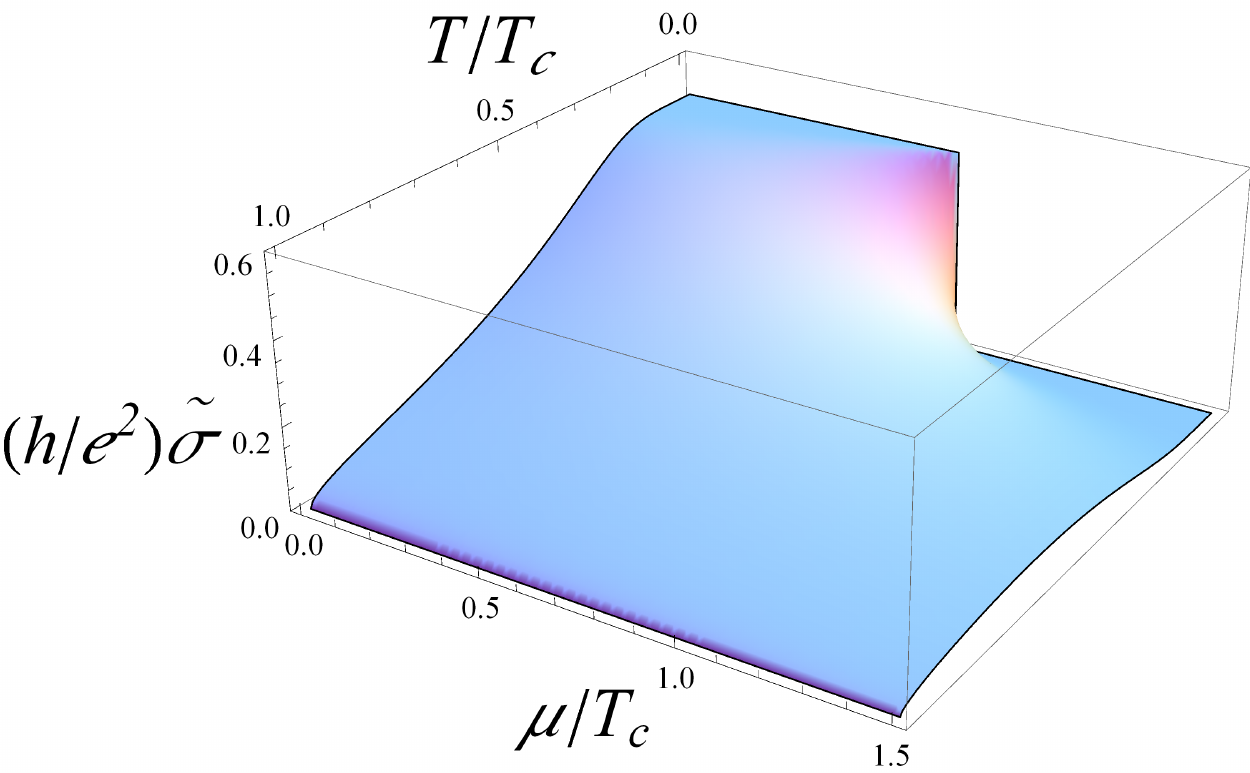}}
~
\subfigure[]{
\includegraphics[scale=0.32]{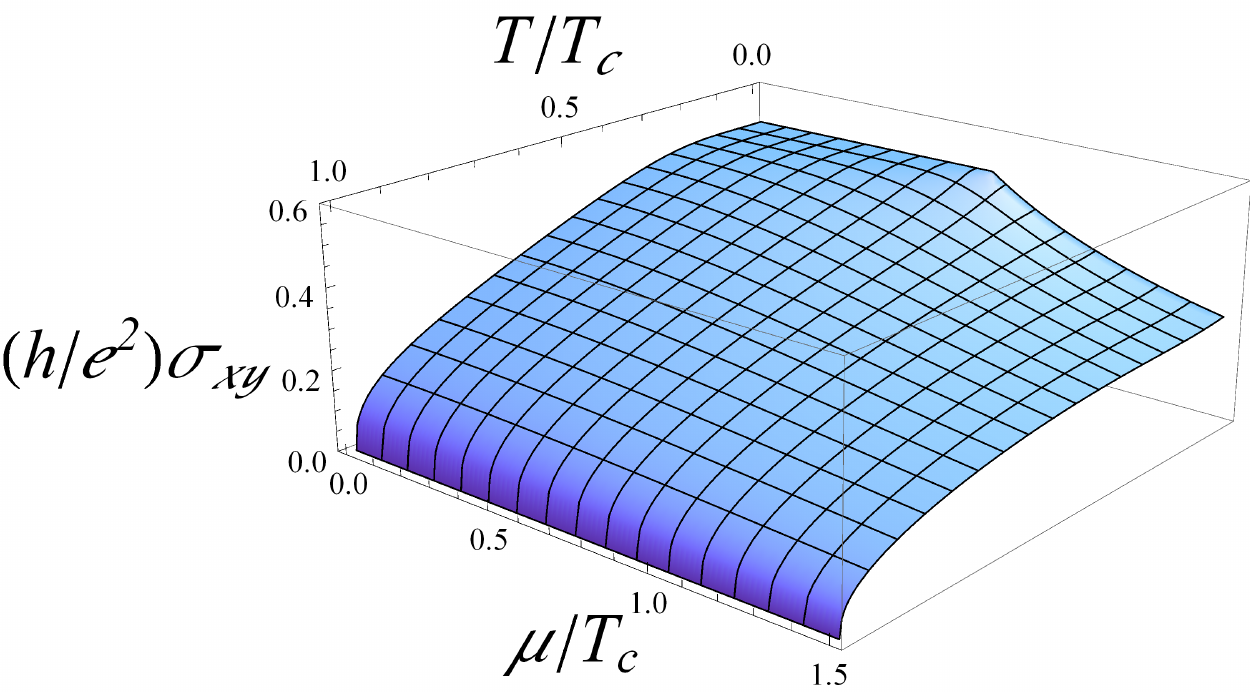}}
 \caption{Calculated topological mass $\tilde \sigma(T,m)$ and Hall conductivity $\sigma_{xy}$
 as a function of temperature, $T$, and the chemical potential, $\mu$, normalized
 by the Curie temperature $T_c$. We consider for $m(T)$ a mean-field temperature dependence,
 $m(T)\approx J_\perp\sqrt{1-T/T_c}$ and use the estimate $J_\perp/T_c\approx 37.4$ [(a) and (b)] based on
 experimental results from Ref. \onlinecite{Moodera} for the Bi$_2$Se$_3$/EuS interface. Note the plateau regions in
 both panels (a) and (b), indicating nearly quantized $\tilde \sigma$ and $\sigma_{xy}$,
 although both $T$ and $\mu$ are finite. There
 is a large region in the $T-\mu$ plane where the Hall conductivity is finite and not quantized, while the topological mass vanishes.
 Panels (a) and (b) depict a situation where $J_\perp$ is large compared to $T_c$. In panels (c) and (d)
 we illustrate a situation with $J_\perp/T_c=0.8$.
 Observe that the region of the plateau is smaller and less sharp. In any situation, the plot of the
 topological mass represents the Hall conductivity only in the regime $\mu<|m|$.}\label{Fig:sigmaT}
\end{figure}
Introducing the unit vector  $\hat d(k)={\bf d}(k)/|{\bf d}(k)|$
associated to the mean-field Hamiltonian ${\cal H}_{\rm MF}={\bf d}(k)\cdot\sigmab=-k_y\sigma_x+k_x\sigma_y+m\sigma_z$, we
can write an expression for $\sigma_{xy}(T,m)$ in terms of the topological charge density,
\begin{equation}
 {\cal Q}_{xy}(k)=\frac{1}{2\pi}\hat d(k)\cdot[\partial_{k_x}\hat d(k)\times\partial_{k_y}\hat d(k)].
\end{equation}
as
%
\begin{equation}
\label{Eq:sigma_xy}
 \sigma_{xy}(T,m)=\frac{e^2}{2h}\int d^2k~{\cal Q}_{xy}(k)
[f_-(k)-f_+(k)],
\end{equation}
where $f_\pm(k)=1/[e^{\beta(\pm|{\bf d}(k)|-\mu)}+1]$ are Fermi-Dirac distributions. Thus, it is straightforward to
compute $\sigma_{xy}$ at $T=0$, \cite{Zang-Nagaosa}
\begin{equation}
\label{Eq:sigma_xy-0}
\sigma_{xy}(0,m_0)=\frac{e^2}{2h}\left[\left({\rm sgn}(m_0)-\frac{m_0}{\mu}\right)\theta(|m_0|-\mu)+\frac{m_0}{\mu}\right],
\end{equation}
and we see that the Hall conductivity at $T=0$
is non-quantized and non-universal in the metallic regime ($\mu>m_0$).
The topological mass, on the other hand, vanishes in the limit $T\to 0$ when $\mu>|m_0|$,
 \begin{equation}
 \label{Eq:spin-conduc-1}
\tilde \sigma(0,m_0)=\frac{e^2}{2h}{\rm sgn}(m_0)\theta(|m_0|-\mu).
\end{equation}
Thus, when the system is in the metallic phase, the topological mass obtained from the low-energy regime of the CS action
does not agree with the Hall conductivity. This means that non-local corrections have to be considered in order to make
the local effective action approach to agree with the expression obtained from linear response.
Further insight on this point is obtained by considering how the topological mass relates to the actual Hall conductivity and see
how it deviates from it in the metallic regime. First, we note that we can write,
\begin{equation}
\label{Eq:spin-conduc}
 \tilde \sigma(T,m)=\sigma_{xy}(T,m)+\tau_{xy}(T,m),
\end{equation}
where
\begin{equation}
\label{Eq:tau_xy}
 \tau_{xy}(T,m)=-\frac{e^2}{2 h}\int d^2k~{\cal Q}_{xy}(k)
 |{\bf d}(k)|\frac{\partial}{\partial\mu}[f_+(k)+f_-(k)],
\end{equation}
The quantity $\tau_{xy}$ yields the deviation of the coefficient of the local contribution to the CS term
from the Hall conductivity when the chemical is above the gap.
%
%
%
We see that for $|m_0|<\mu$
the contribution $\tau_{xy}$ is responsible for canceling the non-quantized contribution at $T=0$, since,
\begin{eqnarray}
 \tau_{xy}(T=0)&=&-\frac{e^2}{2h}\int d^2kQ_{xy}(k)|{\bf d}(k)|\delta(|{\bf d}(k)|-\mu)
 \nonumber\\
 &=&-\frac{e^2m_0}{2h\mu}=-\sigma_{xy}(T=0,|m_0|<\mu),
\end{eqnarray}
precisely canceling the last term in Eq. (\ref{Eq:sigma_xy-0}).
Thus, we see that the chemical potential acts as a cutoff setting the limit of validity of the local effective action.

In Fig. \ref{Fig:sigmaT}(a) we show $\tilde \sigma(T,m)$ for a
typical mean-field theory dependence with the temperature for $m$. In order to plot $\tilde \sigma$ and $\sigma_{xy}$ we have used
values for $J_\perp$ and Curie temperature estimated from the experiment by Wei {\it et al.} \cite{Moodera} performed on Bi$_2$Se$_3$/EuS
samples. There the estimated value for $J_\perp$ at the interface is considerably larger ($\sim 150$ meV \cite{private})
than the one obtained in {\it ab initio} calculations
for Bi$_2$Se$_3$/MnSe,\cite{Qi-2013,Chulkov}
which features an antiferromagnet material (MnSe) rather than a ferromagnetic one.

The numerical
calculation of the Hall conductivity is shown in Fig. \ref{Fig:sigmaT}(b). We note a plateau similar to the one
arising for $\tilde \sigma$, which indicates that quantization nearly holds for finite $T$ and $\mu$ in a region of
the $T-\mu$ plane determined by the inequality $\mu<|m(T)|$.
However, there is also a region in the $T-\mu$ plane where the Hall conductivity is non-quantized while
the topological mass $\tilde \sigma$ vanishes.

Note that sharp and large plateau regions up to $T=T_c$ [Fig. 1, panels (a) and (b)]
occur only if the gap is larger than $T_c$, which is precisely the
case in the experiment from Ref. \onlinecite{Moodera}.
In panels (c) and (d) of Fig. 1 we also illustrate the opposite situation,
taking for example,
$J_\perp/T_c=0.8$. Observe that the plateau region here is not as sharp and loses its approximate quantization long before
$T_c$ is reached.

\section{Magnetization dynamics}

Let us now derive the LL equation in the insulating regime where $\mu<|m|$. In this case
the CS action (\ref{Eq:CS-action}) contributes to the LL equation in two different ways.
This can be see by decomposing
Eq. (\ref{Eq:CS-action}) into two parts,
${\cal S}_{\rm CS}={\cal S}_{\rm Berry}+{\cal S}_{\rm TME}$,
where
\begin{equation}
\label{Eq:Berry}
 {\cal S}_{\rm Berry}=
 \frac{\sigma(T,m)}{2}
 \int dt\int d^2r~({\bf n}\times\hat {\bf z})\cdot
 \partial_t{\bf n},
\end{equation}
is the correction to the Berry phase, and
\begin{equation}
\label{Eq:TME}
 {\cal S}_{\rm TME}=-\frac{e\sigma(T,m)}{J}
 \int dt\int d^2r~{\bf n}\cdot{\bf E},
\end{equation}
is the  magnetoelectric contribution.
Therefore, these two contributions lead to the LL equation,
\begin{equation}
 \label{Eq:LL-main}
 \left[1-\frac{\sigma(T,m)}{2}({\bf n}\cdot\hat {\bf z})\right]\partial_t{\bf n}=
 {\bf n}\times{\bf H}_{\rm eff}-\alpha({\bf n}\times\partial_t{\bf n})+{\bf T}_{\rm TME},
\end{equation}
where
\begin{equation}
{\bf T}_{\rm TME}=\frac{e\sigma(T,m)}{J}{\bf n}\times{\bf E}
\end{equation}
is the magnetoelectric torque.
Note that within the one-loop accuracy of our calculations
$(\hat {\bf z}\cdot{\bf n})\partial_t{\bf n}\approx\langle n_z\rangle\partial_t{\bf n}=(m/J_\perp)\partial_t{\bf n}$.
%
Thus, due to the magnetoelectric torque, the spin-wave excitation will be gapped in general. For instance, if
${\bf E}=-E\hat {\bf x}$ with $E={\rm const}$, we have,
\begin{equation}
 \omega_{\rm sw}(k)=\frac{1}{1-\frac{m\sigma}{2J_\perp}}\sqrt{H_{\rm eff}^2(k)+\frac{e^2\sigma^2}{J^2}E^2},
\end{equation}
where in the absence of an external magnetic field, $H_{\rm eff}(k)\to 0$ as $k\to 0$.

\section{Conclusions}

We have discussed the screening of the Coulomb interaction on the surface of a TI at finite temperature and
chemical potential. In the presence of proximity-induced ferromagnetism we found a peculiar behavior where
the screening length depends on the magnetization via the electronic gap. This implies in particular that no screening
occurs at zero temperature if
the chemical potential is smaller than the proximity-induced fermionic gap.
We have also calculated the coefficient of the CS term (the topological mass) analytically for finite temperature and chemical potential in
the regime $\mu<|m|$. This topological mass yields the Hall conductivity in this regime and
directly influence the magnetization dynamics at the
surface, modifying the spin-wave velocity and inducing an magnetoelectric gap in the spin-wave excitation spectrum.
Interestingly, in the insulating regime the topological mass remains nearly quantized even at finite temperature and chemical
potential.

\acknowledgments

We would like to thank the SFB TR 12 program of the Deutsche Forschungsgemeinschaft (DFG) for the financial
support. The work of IE is also supported by the Ministry of Education and Science of
the Russian Federation in the framework of Increase Competitiveness
Program of NUST “MISiS” (Nr. K2-2014-015)

\appendix
\section{}

In this appendix we will discuss the integral (\ref{Eq:Integral}) in more detail. Let us first consider
the case where $\omega_n=0$ and ${\bf p}=0$ simultaneously. In this case it is very simple to perform the momentum integral to
obtain,
\begin{eqnarray}
\label{Eq:I00}
 I(0,0)&=&-\frac{T}{4\pi}\sum_n\frac{1}{(i\nu_n+\mu)^2-m^2}
 \nonumber\\
 &=&\frac{1}{8\pi|m|}[f_-(0)-f_+(0)],
\end{eqnarray}
which after some straightforward simplifications lead to Eq. (\ref{Eq:sigma}).

Now, let us set ${\bf p}=0$ and leave the bosonic Matsubara frequency $\omega_n\neq 0$. If we now perform the fermionic
Matsubara sum, we obtain,
\begin{equation}
\label{Eq:Iw0}
 I(\omega_n,0)=\int\frac{d^2k}{(2\pi)^2}\frac{f_+(k)-f_-(k)}{E(k)[(i\omega_n)^2-4E^2(k)]},
\end{equation}
where $E(k)=\sqrt{k^2+m^2}$. If we now take the limit $\omega_n\to 0$, we obtain
the result needed to compute the Hall conductivity. Clearly, $\lim_{\omega_n\to 0}I(\omega_n,0)$ is not the same as $I(0,0)$
in Eq. (\ref{Eq:I00}). Thus, the Matsubara sum and momentum integral do not commute with the limit $\omega_n\to 0$. We note that
when $\omega_n\to 0$ before performing the Matsubara sum, the poles coalesce, such that an additional contribution arises, which
makes $\tilde \sigma$ to differ from the Hall conductivity by the term $\tau_{xy}$ [recall Eq. (\ref{Eq:spin-conduc})].

Just to compare further with Eq. (\ref{Eq:sigma_xy}), note that $|{\bf d}(k)|=E(k)$, and
\begin{equation}
 {\cal Q}_{xy}(k)=\frac{1}{2\pi}\hat d\cdot(\partial_{k_x}\hat d\times\partial_{k_y}\hat d)=\frac{m}{2\pi E^3(k)},
\end{equation}
such that,
\begin{equation}
 \lim_{\omega_n\to 0}I(\omega_n,0)=\frac{1}{8\pi |m|}\int d^2k{\cal Q}_{xy}(k)[f_-(k)-f_+(k)].
\end{equation}

\end{document}